\documentclass[usenatbib, useAMS]{mnras}

\usepackage{bm}

\usepackage{graphicx}

\newcommand{\arepo}{{\sc arepo}}

\newcommand{\kms} {{\rm km~s}^{-1}}

\newcommand{\Mpc} {{\rm Mpc}}
\newcommand{\mo}{{\rm M}_{\sun}}

\newcommand{\gsim}{\lower.7ex\hbox{$\;\stackrel{\textstyle>}{\sim}\;$}}
\newcommand{\lsim}{\lower.7ex\hbox{$\;\stackrel{\textstyle<}{\sim}\;$}}
\newcommand{\sgsim}{\lower.6ex\hbox{$\;\stackrel{\scriptstyle>}{\scriptstyle\sim}\;$}}
\newcommand{\slsim}{\lower.6ex\hbox{$\;\stackrel{\scriptstyle<}{\scriptstyle\sim}\;$}}

\newcommand{\G}{{\rm G}}
\newcommand{\nG}{{\rm nG}}
\newcommand{\muG}{{\rm \mu G}}

\newcommand{\rev}[1]{#1}

\defcitealias{Planck2015}{Planck Collaboration XIX} 
\defcitealias{Neronov2010}{Neronov \& Vovk 2010} 

\newcommand{\planckpap}{\citetalias{Planck2015} \citeyearpar{Planck2015}}
\newcommand{\planckpappar}{\citetalias{Planck2015} \citeyear{Planck2015}}

\begin{document}
\title[B field effects on the galaxy population]
{Effects of simulated cosmological magnetic fields on the galaxy population} 
\author[Marinacci \& Vogelsberger]
{Federico Marinacci\thanks{E-mail:
fmarinac@mit.edu} and Mark Vogelsberger\vspace*{0.2cm}\\
  Kavli Institute for Astrophysics and Space Research, 
  Massachusetts Institute of Technology, Cambridge, MA 02139, USA}

\date{Accepted 2015 November 7.  Received 2015 October 17; in original form 2015 August 26.}  

\pagerange{\pageref{firstpage}--\pageref{lastpage}}
\pubyear{2015}

\maketitle

\label{firstpage}

\begin{abstract}
We investigate the effects of varying the intensity of the primordial magnetic 
seed field on the global properties of the galaxy population in ideal magnetohydrodynamic
cosmological simulations performed with the moving-mesh code \arepo. We vary the 
seed field in our calculations in a range of values still compatible with the 
current cosmological upper limits. We show that above a critical intensity of 
$\simeq 10^{-9}\,\G$, the additional pressure arising from the field strongly 
affects the evolution of gaseous structures, leading to a suppression of the 
cosmic star formation history, \rev{which is stronger for larger seed 
fields. This directly reflects into a lower total galaxy count above a
fixed stellar mass threshold at all redshifts, and} a lower galaxy number density at fixed 
stellar mass and a less massive stellar component at fixed virial mass at all 
mass scales. These signatures may be used, in addition to the existing methods, 
to derive tighter constraints on primordial magnetic seed field intensities. 
\end{abstract}


\section{Introduction}\label{sec:intro}
Much progress has been made to understand the evolution of magnetic fields in 
the cosmological context, and especially on investigating mechanisms 
through which B fields are set to their present-day strength in galaxies and 
galaxy clusters. Although there are strong indications that dynamo processes 
play a crucial role in amplifying pre-existing B fields \citep{Hanasz2009, 
Schober2013}, the origin of these seed fields still remains uncertain. In 
particular, it is unclear whether these seeds are of cosmological origin, i.e. 
produced at shocks \citep{Ryu1998} or during inflation and phase transitions 
in the early Universe \citep{Widrow2012}, or generated by 
stars in proto-galaxies \citep{Schleicher2010} and subsequently ejected by 
galactic winds \citep{Volk2000} and active galactic nucleus activity \citep{Furlanetto2001}.

It is intrinsically difficult to observe magnetic fields outside haloes or at
high redshift with the current generation of instruments
\citep{Beck2007}. The detection of magnetic fields outside collapsed structures
is particularly challenging because the expected intensities ($\lsim
1\,\nG$) are well below the ones observed in galaxies and galaxy clusters
($\sim 1\,\muG$). Hence, most of the constraints on primordial magnetic
fields are upper limits. These are in turn derived by considering how the
predicted effect of such seed fields would compare to actual observations of more
well-determined quantities \rev{such as the cosmic microwave background power 
spectrum or polarization 
(\citeauthor{Barrow1997} \citeyear{Barrow1997};
\citeauthor{Durrer2000} \citeyear{Durrer2000};
\citeauthor{Jedamzik2000} \citeyear{Jedamzik2000};
\citeauthor{Kahniashvili2009} \citeyear{Kahniashvili2009}, \citeyear{Kahniashvili2010};
\citetalias{Planck2015} \citeyear{Planck2015}), the isotropy in the distribution
of ultrahigh energy cosmic rays \citep[e.g.][]{Lee1995}, Big bang nucleosynthesis
abundances \citep{Grasso1995,Grasso2001}, Fardaday rotation measure, Ly $\alpha$ forest
and Sunyaev--Zel'dovich effect statistics \citep{Blasi1999, Shaw2012}.}
Interestingly, most of the limits converge towards a maximum present-day field
strength of $\sim 1-10\,\,\nG$. \rev{Moreover, possible \textit{lower}
limits ($\sgsim 10^{-15}\,\G$) have been inferred from $\gamma$-ray 
and TeV blazars observations \citep{Neronov2009, Neronov2010}}.

In this Letter, we examine how strong the effects of cosmological B fields (i.e. 
fields generated in the early Universe) on the global properties of the galaxy 
population are as a function of the field strength, and determine what is the 
critical intensity at which they become noticeable. We run a set of cosmological 
magnetohydrodynamic (MHD) simulations with varying seed field intensities and study the changes in the 
\rev{cosmic} star formation density, stellar mass function, stellar-to-halo mass 
relation, the evolution of the mean magnetic field \rev{and total galaxy 
number}. The Letter is organized as follows. In Section~\ref{sec:analytic}, we 
present analytic considerations to relate the seed field used in our runs to the 
current upper limits and we estimate at what intensity those fields are expected 
to significantly impact the dynamics of the gas. We introduce the setup of our 
simulations in Section~\ref{sec:method} and we describe the main results of our 
analysis in Section~\ref{sec:results}. Finally, we draw our conclusions in  
Section~\ref{sec:conclusions}.

\section{Analytic considerations}\label{sec:analytic}

\rev{In our simulations, the initial seed magnetic field is 
uniform and threads the computational domain along a predefined direction
(see also Sect. \ref{sec:method}). However, in the majority of earlier work,
upper limits for cosmological magnetic fields are inferred by assuming that 
those fields are described by a power spectrum of the form }
\citep[see e.g.][]{Kahniashvili2009}
\begin{equation}
 P_{\rm B}(k) = A_{\rm B} k^{n_{\rm B}}\,\,\,\ k < k_{\rm D},
 \label{eq:spectrum}
\end{equation}
where $A_{\rm B}$ is a normalization constant, $n_{\rm B}$ is the spectral index and $k_{\rm D}$ 
is the so-called Alfv\'en damping scale \citep{Jedamzik1998} \rev{above} which 
it is assumed that no power is present. \rev{Therefore, it is important to establish 
a direct connection between the uniform field intensities used in this work and those 
derived through equation (\ref{eq:spectrum}). Before proceeding, we note 
that the value of $n_{\rm B}$ is uncertain (usually $-3< n_{\rm B} \leq 2$; 
\citealt{Kahniashvili2010}) and can have a significant impact 
on the value of the inferred intensities of cosmological B fields. It is
also to avoid this additional source of uncertainty  that we opted for 
the simpler uniform field setup in our runs}.

\rev{The information about the intensity of cosmological seed B fields
is contained in the normalization constant $A_{\rm B}$, which is linked 
to a smoothed comoving field amplitude on a comoving scale 
$\lambda$ via a convolution with a Gaussian
kernel through (\planckpappar)}
\begin{equation}
 B^2_{\lambda} = \int_{0}^{+\infty} \frac{{\rm d}k\,k^2}{2\pi^2}e^{-k^2\lambda^2}\,P_{\rm B}(k) = 
 \frac{|A_{\rm B}|}{4\pi^2\lambda^{n_{\rm B}+3}}\Gamma\left(\frac{n_{\rm B}+3}{2}\right),
 \label{eq:normalization}
\end{equation}
\rev{where $\Gamma$ is Euler's gamma function.}
From equation~(\ref{eq:normalization}) it is easy to connect the smoothed 
field amplitude at two arbitrary scales $\lambda_1$ and $\lambda_2$ as
\begin{equation}
 B_{\lambda_2} = B_{\lambda_1}\left(\frac{\lambda_1}{\lambda_2}\right)^{\frac{n_{\rm B}+3}{2}}.
\end{equation}
In our simulation setup, the initial seed field is coherent on a $25\,{\rm h^{-1}}\Mpc 
\simeq 37~\Mpc$ scale. \rev{Hence}, a smoothed field $B_{1\,{\rm 
Mpc}} = 1\,\nG$ -- a representative value of current upper limits 
(\citetalias{Planck2015} \citeyear{Planck2015}) -- translates 
into $B_{37\,{\rm Mpc}} \simeq 1.2\times10^{-4}\,\nG$ for $n_{\rm B} = 2$, 
and $B_{37\,{\rm Mpc}} \simeq 0.84\,\nG$ for $n_{\rm B} = -2.9$.

A more physically motivated way to express the intensity of a magnetic seed 
field whose power spectrum satisfies equation (\ref{eq:spectrum}) is to replace 
$B_{\lambda}$ with the \rev{\textit{effective}} field containing the same magnetic 
energy density as the one obtained by integrating the power spectrum over all 
scales. \rev{In this way, the information at all spatial scales is used and the 
resulting field intensity can be compared in a more meaningful way to the
uniform seed field that we have adopted in our calculations}. 
\rev{The effective} field is defined as \citep[][equation 5]{Kahniashvili2010} 
\begin{equation}
B_{\rm eff} = \frac{B_\lambda(k_{\rm D}\lambda)^{(n_{\rm B}+3)/2}}{\Gamma^{1/2}(n_{\rm B}/2 + 5/2)}.
\label{eq:Beff}
\end{equation}
\rev{By applying equation~(\ref{eq:Beff})} in the case of the $n_B$ interval 
that we have examined above, we obtain an effective field $B_{\rm eff} \simeq 
1.22\,\nG$ and $B_{\rm eff} \simeq 240\,\nG$ for $n_{\rm B} = -2.9$ and $n_{\rm B} = 2$, 
respectively. These values have been derived for $B_{1 \rm Mpc} = 1\,\nG$ and by 
adopting the expression for $k_{\rm D}$ given in equation (3) of \planckpap. The maximum 
seed field value that we employ in our simulations is consistent with those 
limits.

\rev{As we mentioned in the Introduction}, the majority of \rev{the currently available upper limits} translate into 
effective seed fields with intensities of the order of $1-10\,\nG$. We now 
present two simple arguments showing that significant effects on the evolution 
of the galaxy population must be expected if the Universe is permeated by those 
(or stronger) fields. We start by noting that in ideal MHD, the magnetic 
field flux is conserved. This leads to the scaling relation \citep{Grasso2001} 
\begin{equation}
 ||\bm{B}_{\rm gal}|| = 10 \left(\frac{||\bm{B}_0||}{1\,\nG}\right) \left(\frac{\delta}{10^6}\right)^{2/3} \muG,
 \label{eq:nodynamo}
\end{equation}
where $\bm{B}_0$ is the intensity of the initial seed field and $\delta$ is 
the density contrast between the average gas density within galaxies and that in 
the intergalactic space. Equation~(\ref{eq:nodynamo}) states that the final field 
intensity observed nowadays in galaxies ($\sim$ a few $\muG$) can be entirely 
accounted for by adiabatic compression of an initial seed field of $1\,\nG$. 
However, several numerical studies \citep[see][and 
references therein]{Marinacci2015} have shown that additional processes, 
such as small-scale dynamo and turbulent or shearing flows, \rev{play a key
role in amplifying} the magnetic field to 
its final strength. \rev{It is thus plausible} to assume that seed fields of 
$\sim 1\, \nG$ would be over-amplified by the presence of these (small-scale) 
mechanisms. This over-amplification would in turn leave its imprint on galactic 
properties by enhancing the dynamical role of magnetic fields.

Magnetic fields start to play a role in the gas dynamics if the magnetic 
pressure becomes comparable to the thermal gas pressure, viz. 
\begin{equation}
 ||\bm{B}||^{2} \gsim 8\pi \rho \frac{kT}{\mu m_{\rm p}},
 \label{eq:Bpress}
\end{equation}
where $||\bm{B}||$ is the magnetic field intensity $\rho$ and $T$ the gas 
density and temperature, $k$ the Boltzmann constant, $\mu$ the gas mean molecular 
weight and $m_{\rm p}$ the proton mass. By considering a fully-ionized gas of
primordial composition ($X = 0.76$), we can rewrite equation (\ref{eq:Bpress}) in terms 
of the comoving magnetic field intensity as 
\begin{equation}
||\bm{B}|| \gsim 5.61\left(\frac{\Omega_{\rm b}h^2}{0.022}\right)^{1/2} \!\!\!\!
 \left(\frac{T}{10^4\,{\rm K}}\right)^{1/2} \!\!(1+z)^{-1/2}\; \nG,
\label{eq:Bcritical}
\end{equation}
where $h = H_0 / 100\,\kms\,\Mpc^{-1}$, $\Omega_{\rm b}$ is the baryon density and $z$ 
the redshift.

\begin{figure*}
\begin{minipage}[c]{\textwidth}
\centering
\ifpdf
\resizebox{0.32\textwidth}{!}{\includegraphics[width=0.25\textwidth]{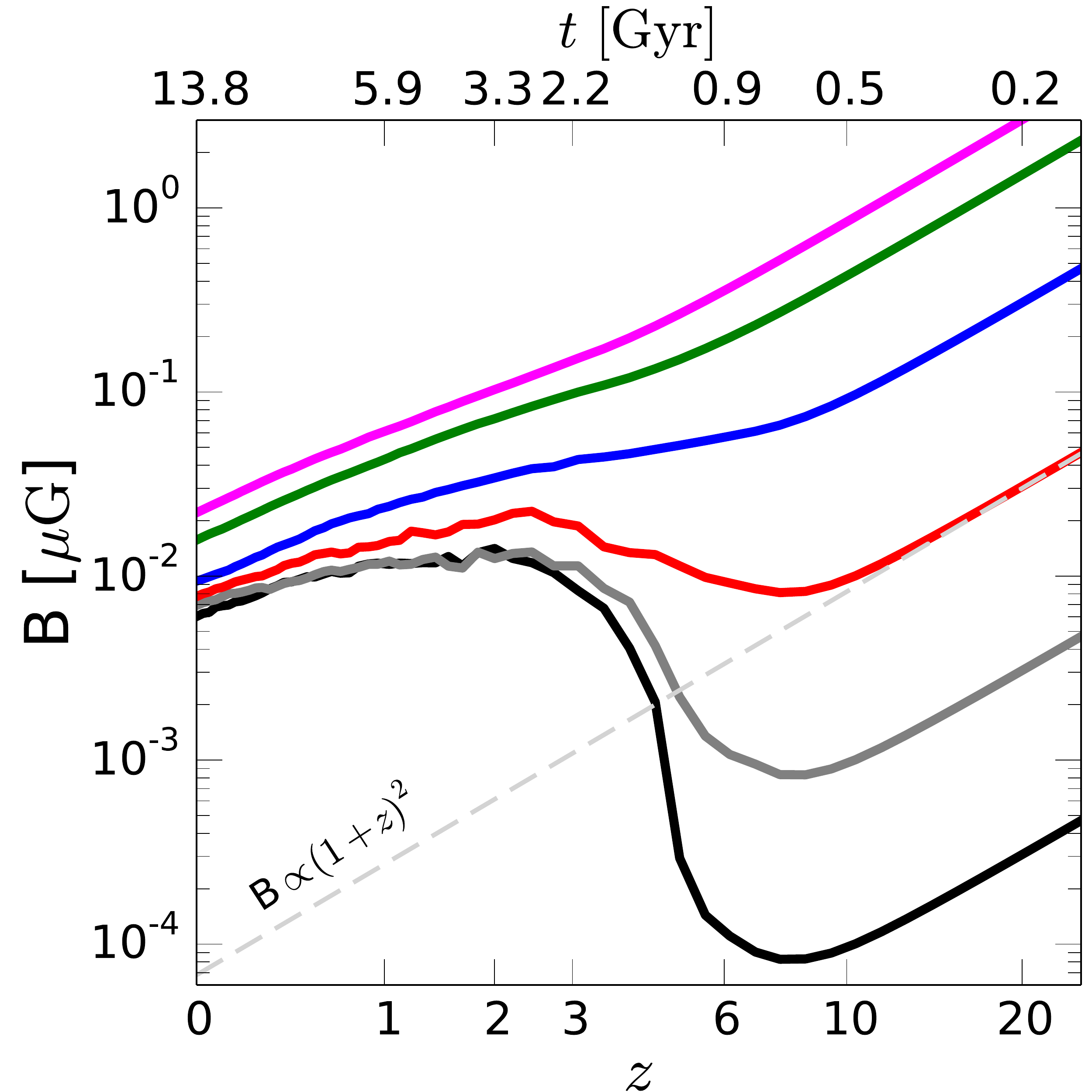}}
\resizebox{0.32\textwidth}{!}{\includegraphics[width=0.25\textwidth]{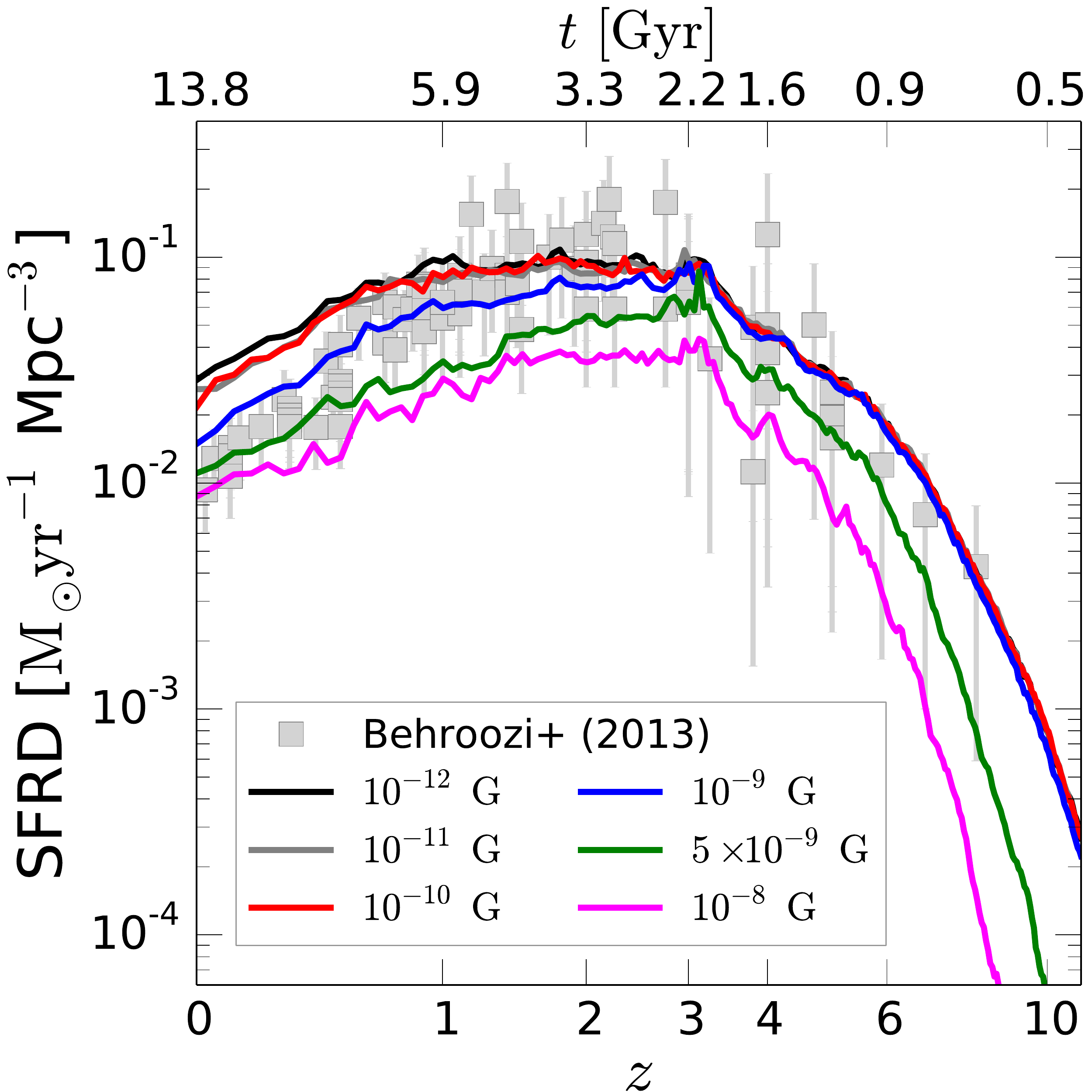}}
\resizebox{0.32\textwidth}{!}{\includegraphics[width=0.25\textwidth]{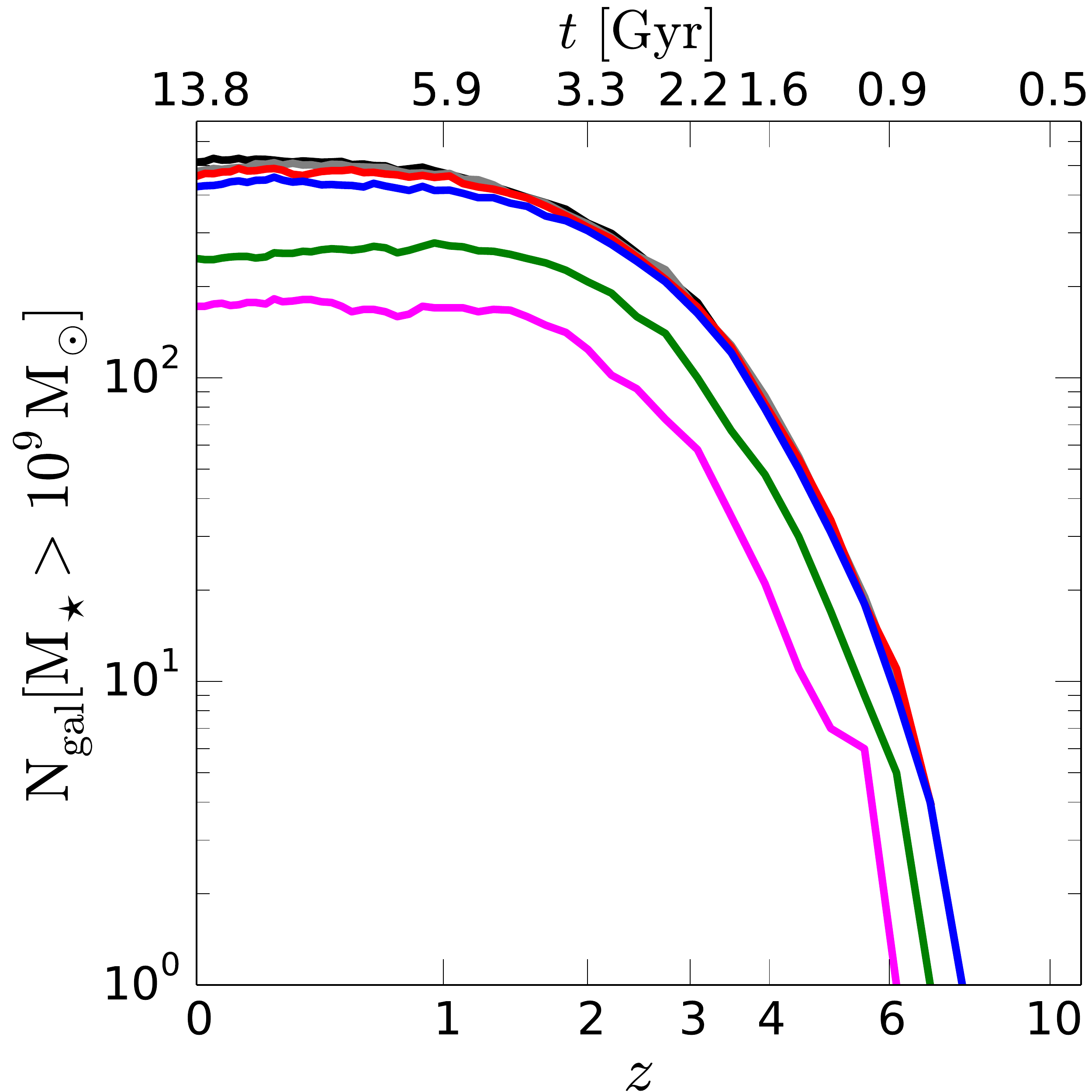}}
\else
\resizebox{0.32\textwidth}{!}{\includegraphics[width=0.25\textwidth]{fig1a.eps}}
\resizebox{0.32\textwidth}{!}{\includegraphics[width=0.25\textwidth]{fig1b.eps}}
\resizebox{0.32\textwidth}{!}{\includegraphics[width=0.25\textwidth]{fig1c.eps}}
\fi
\end{minipage}
\caption{Left: evolution of the volume-weighted rms B field 
strength. Above the critical B field value of $10^{-9}\,\G$ the field intensity 
is set by flux conservation and little turbulent amplification is present. 
Centre: evolution of the cosmic star formation rate density (SFRD) 
together with  a compilation of observational data taken from 
\citet{Behroozi2013}. For larger seed magnetic field strengths there is an 
increasing suppression of the SFRD, which is present at all times for seed 
fields $>10^{-9}\,\G$. \rev{Right: evolution of the total galaxy number 
above a stellar mass threshold of $10^9\mo$. The decrease in the total number
of objects for seed field $>10^{-9}\,\G$ is visible at all times}. } 
\label{fig:prop}
\end{figure*}

Equation (\ref{eq:Bcritical}) shows that the transition to magnetically
dominated gas flows occurs at a 
critical seed magnetic field strength between $1$ and $10\,\,\nG$, in line with the estimates 
and \rev{the observational constraints discussed earlier}. In 
particular, in our runs the gas reaches temperatures of $\simeq 10^4~{\rm K}$ 
at the end of the reionization, which occurs at $z_{\rm reion} = 6$ 
for our choice of the UV background \citep{FaucherGiguere2009}. This implies 
a critical seed B field $||\bm{B}|| \gsim 1.44\;\nG$. 
Therefore, we 
anticipate a significant effect on galactic evolutionary processes at these B 
field intensities.

\section{Simulations setup}\label{sec:method}
Our simulations evolve a uniformly sampled box from the 
initial redshift $z=127$ to the present day. \rev{We adopt a $\Lambda$ cold dark 
matter cosmology with parameters $\Omega_{\rm m} = 0.302$, 
$\Omega_{\rm b} = 0.04751$, $\Omega_{\rm \Lambda} = 0.698$, $\sigma_{8} = 
0.817$, $n_{\rm s} = 0.9671$, and $H_{0} = 68~\kms\Mpc^{-1}$ derived by an 
independent re-analysis of {\it Planck} data by \cite{Spergel2015}}. \rev{All 
the runs} are performed with the moving-mesh code \arepo\ \citep{Arepo}, 
together with a module for ideal MHD \citep{Pakmor2013} based on the 8-wave 
\cite{Powell1999} formalism to control divergence errors. The code is 
complemented with a galaxy formation physics model developed for the {\sc 
illustris} simulation suite \citep{Vogelsberger2014}, \rev{which includes as 
major components: (1) primordial and metal-line cooling with self-shielding 
corrections \citep{Rahmati2013}, (2) an updated version of the 
\citet{SFR_paper} model for star formation with a \citet{Chabrier2003} initial 
mass function, (3) stellar evolution, gas recycling and chemical enrichment, 
(4) a kinetic galactic wind scheme in which the wind speed is determined by the 
local dark matter velocity dispersion in haloes \citep[see e.g.][]
{Puchwein2013}, (5) black hole seeding, accretion, merging and feedback 
\citep[see also][]{BH_paper}}. In this work we use the fiducial 
settings of the model presented in \cite{Vogelsberger2013}. 

We sample a 
$25~{\rm h^{-1}}\Mpc$ box size with $2\times 256^3$ resolution elements. Magnetic 
fields are added at the beginning of the calculation as a uniform seed field 
throughout the box. \rev{We choose this initial setup because it is the 
simplest divergence-free configuration to implement and has also the advantage that 
the field is specified once its intensity is fixed \citep[but see][for a 
discussion on how the initial field direction affects its 
properties]{Marinacci2015}.} We keep the direction of the field fixed 
in all simulations \rev{along the $z$-axis}, but change the (comoving) strength 
of the initial seed in the interval $[10^{-12} - 10^{-8}]\,\,\G$.

\section{Results}\label{sec:results}
The \rev{left-hand} panel of Fig.~\ref{fig:prop} presents the time evolution of the 
(volume-weighted) rms B field intensity as a function of redshift. The initial 
seed field intensity is indicated in the legend. At low seed fields ($< 
10^{-9}\,\G$), after a first phase dominated by the expansion of the Universe, 
\rev{in which the B field strength declines $\propto(1+z)^2$ (grey dashed 
line)}, there is an upturn of the mean B field intensity at \rev{$z \approx 6$}, after 
which the field is amplified \rev{by structure formation}. The amplification 
saturates quickly, leaving behind an average field of $\sim 10^{-2}\,\muG$, 
whose amplitude slowly declines with time. The situation is different for seed 
fields $\geq 10^{-9}\,\G$. In those cases, the amplification of the field 
\rev{due to build-up of cosmic structures} is \rev{less prominent}. Indeed, 
the magnetic field strength steadily drops with time and adiabatic compression 
alone is largely sufficient to set the B field intensities observed in the 
present-day Universe (see eq.~[\ref{eq:nodynamo}]). 

The increased magnetic field strength for large seed fields implies a larger 
magnetic pressure acting on the gas, which in turn provides more support against 
gravitational collapse, potentially affecting the efficiency of its conversion 
into stars. We illustrate this in the \rev{central} panel of Fig.~\ref{fig:prop} in 
which we show the cosmic star formation rate density (SFRD) as a function of 
redshift compared to a compilation of observational determinations taken from 
\citet{Behroozi2013}. We can roughly divide the simulation set in two groups, 
depending on whether the initial seed field is below or above the critical value 
of $10^{-9}\,\G$, which is very close to the one that we have determined 
analytically. Below this value, the magnetic field is still dynamically 
negligible (see eq.~[\ref{eq:Bcritical}]) and the effects on the SFRD are 
minimal. Conversely, seed fields $\geq 10^{-9}\,\G$ have a significant impact on 
gas dynamics. The result is a noticeable reduction of the global SFRD, which is 
more prominent the larger the seed field and increasingly falls short of the 
observational findings especially at the star formation peak at $z\sim2$, while 
reconciling the simulations to the observed level of star formation at 
$z\simeq0$. The simulation with seed field $10^{-9}\,\G$ exhibits an intermediate 
behaviour. At $z\gsim3$, it follows very closely the same trend of the 
simulations with a lower seed field. A reduction in the SFRD (about a factor of 
2 with respect to the runs with seed field $<10^{-9}\,\G$) only occurs after 
$z\sim3$.

\rev{The right-hand panel of Fig.~\ref{fig:prop} displays the evolution of the 
total number of galaxy above a stellar mass threshold of $10^9\mo$. The stellar
mass is determined as the mass in stars contained within \textit{twice} 
the stellar half-mass radius of each object. From the plot, it can be readily
appreciated that the number of galaxies grows very rapidly for $z \lsim 6$ and
by $z \simeq 2$ (approximately at the peak of the cosmic SFRD) already a 
significant fraction ($\gsim 60\%$) of the total objects has formed. The 
suppression of the cosmic SFRD due to increasing B seed fields intensities
translates into a decrease of the total number of galaxy formed. The decrease is
more pronounced for larger B field strengths and particularly noticeable for 
all the runs with fields above the critical intensities value of $10^{-9}\,\G$. In
those cases, the deficiency in the total galaxy number relative to simulations
with seed field strengths below the critical value -- which closely track one another 
-- is present at all redshifts. Again, the $10^{-9}\,\G$ simulation shows an intermediate
behaviour starting to slightly deviate from the trend followed by lower seed field
runs only after $z \simeq 2$. We explicitly checked that the general trends 
discussed above are roughly independent from the particular value of the 
stellar mass cut adopted, at least up to $10^{10}~\mo$ where minor differences
in the final total number of objects formed start to be present also in runs with 
seed field intensities $< 10^{-9}\,\G$.}

\begin{figure}
\begin{minipage}[c]{0.5\textwidth}
\centering
\ifpdf
\resizebox{0.84\textwidth}{!}{\includegraphics[viewport=15 0 730 680,clip=true,width=0.25\textwidth]{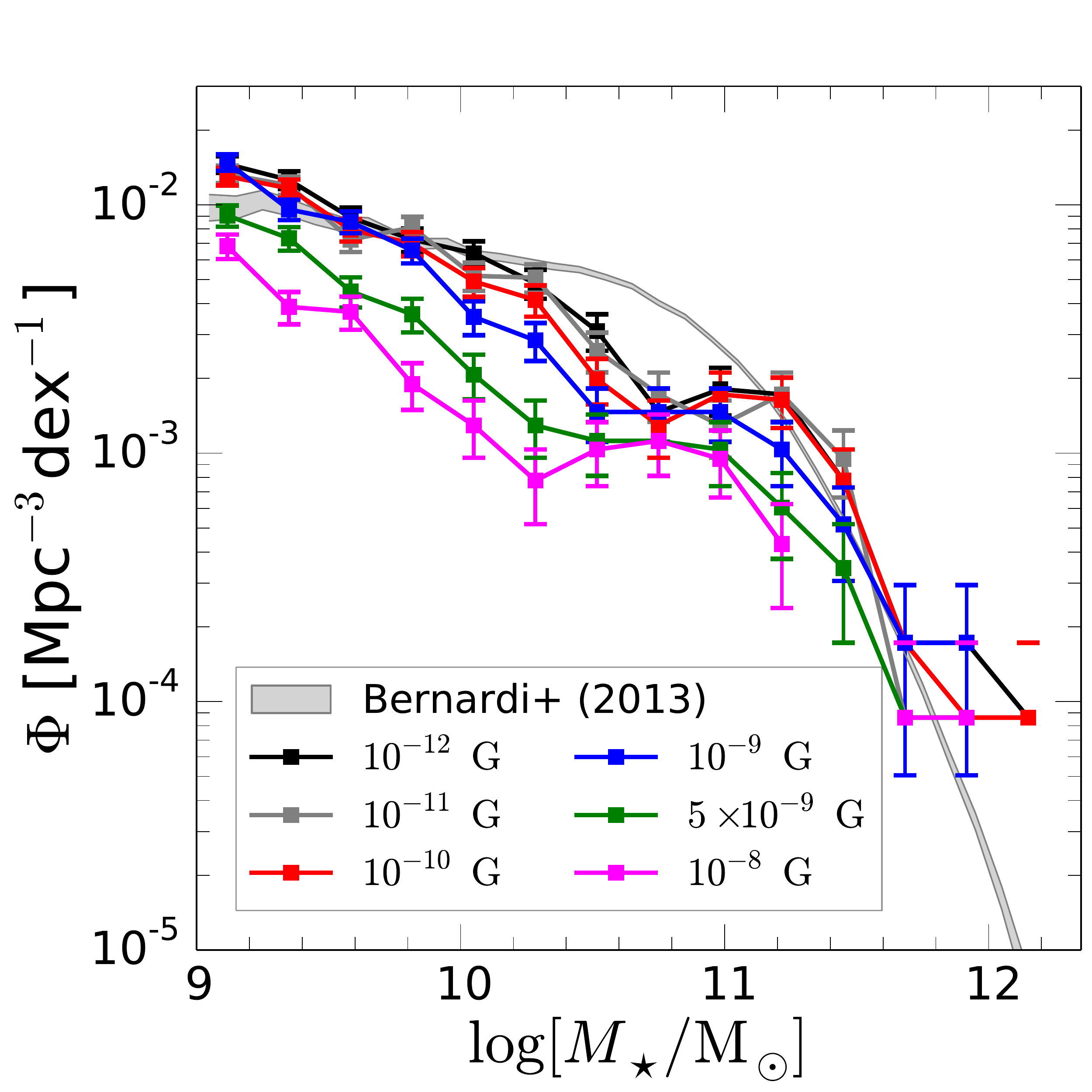}}

\resizebox{0.84\textwidth}{!}{\includegraphics[viewport=15 0 730 680,clip=true,width=0.25\textwidth]{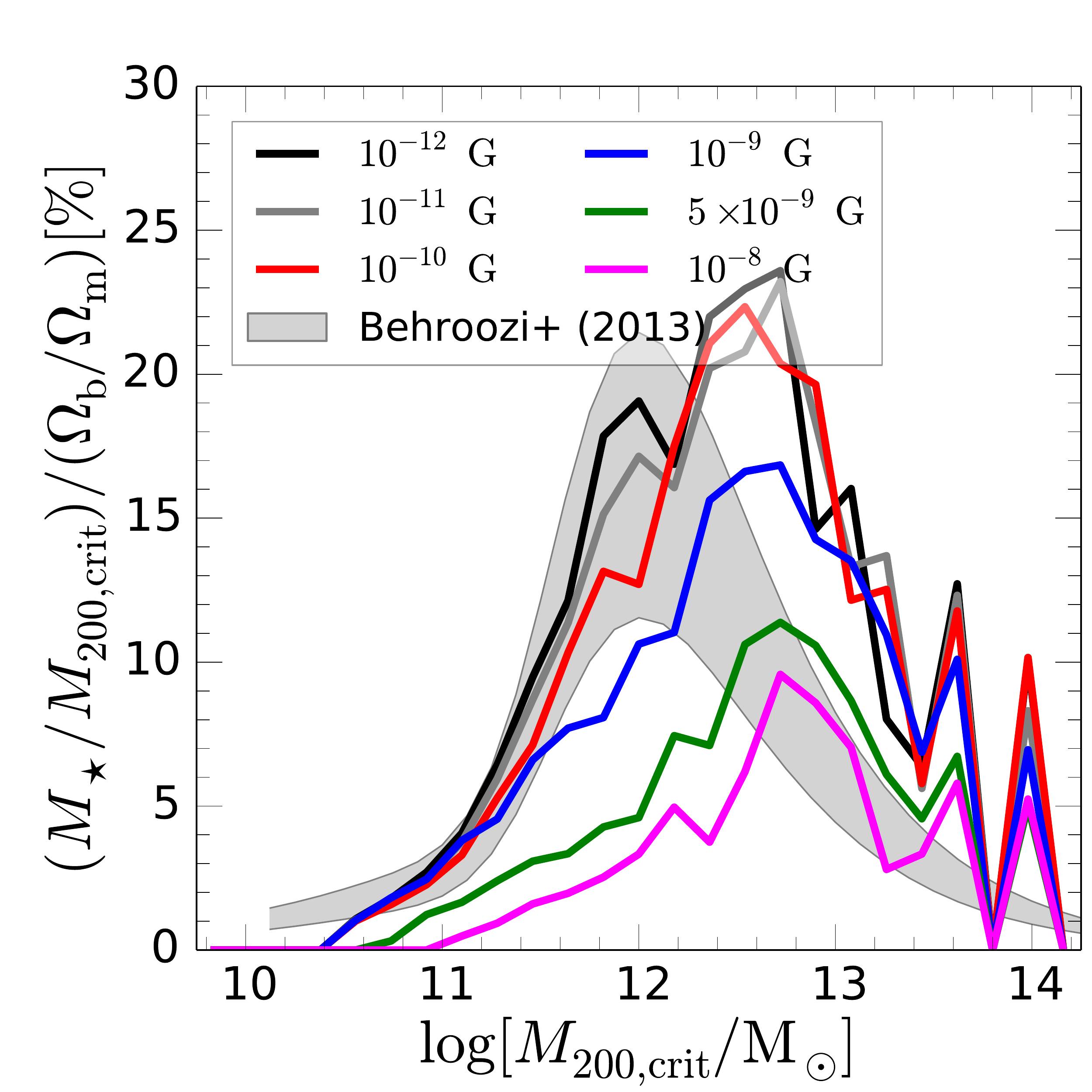}}
\else
\resizebox{0.84\textwidth}{!}{\includegraphics[viewport=15 0 730 680,clip=true,width=0.25\textwidth]{fig2a.eps}}

\resizebox{0.84\textwidth}{!}{\includegraphics[viewport=15 0 730 680,clip=true,width=0.25\textwidth]{fig2b.eps}}
\fi
\end{minipage}
\caption{Top: galaxy stellar mass function 
compared to SDSS observations \citep{Bernardi2013}. The suppression of star 
formation causes a decrease of the number of objects at all mass scales. The 
effect is stronger for larger seed field intensities. Bottom: 
stellar-to-halo mass relation contrasted to the abundance matching predictions 
of \citet{Behroozi2013}. At any given virial mass, the stellar content of haloes 
decreases for increasing seed B field strengths. } 
\label{fig:prop2}
\end{figure}

The \rev{top} panel of Fig.~\ref{fig:prop2} shows the redshift zero galaxy 
stellar mass function for all simulations together with observational values 
inferred from SDSS data \citep{Bernardi2013}. Due to the inability of small 
haloes to accrete highly magnetized gas, we expect a global reduction of the 
number of galaxy per unit stellar mass for increasing seed field intensities. 
The plot fully confirms these expectations and shows that the effect is stronger 
the larger the initial seed field. For seed fields $\leq 10^{-9}\,\G$, the mass 
functions agree, within the scatter, over all the considered mass range. For 
seed fields above $10^{-9}\,\G$, the reduction in number density is clearly 
visible up to $M_\star \sim 10^{10.5}\,\mo$, where the number of galaxies 
is underpredicted compared to the observations. At a stellar mass of $\sim 
10^{10.5}\,\mo$, the discrepancy among the different runs is less prominent, 
an effect likely due to the fact that these haloes are massive enough to power 
their star formation even in the presence of a high B field. For larger 
$M_\star$ (up to $\sim 10^{11.5}\,\mo$) the differences grow again, 
signalling that the building blocks of these large structures, which accrete a 
significant fraction of their mass through mergers, contain on average less 
stars for large seed fields.

In the \rev{bottom} panel of Fig.~\ref{fig:prop2}, we present the redshift zero 
stellar-to-halo mass relation as a function of the virial mass. For comparison,
we also show the same relation derived via abundance matching 
\citep{Behroozi2013}. Again, the net effect of an increase of the seed field 
strength is a reduction of the ratio between stellar and halo mass over the 
whole range of virial masses in the simulations. The general division in two 
groups discussed previously can also be applied here. In particular, simulations 
with seed fields $< 10^{-9}\,\G$ track each other quite well, while for seed 
fields $\geq 10^{-9}\,\G$, the stellar-to-halo mass ratio becomes smaller for 
increasing value of the initial seed magnetic field. At low halo masses, the 
conversion efficiency of gas into stars is considerably reduced for the highest 
seed magnetic fields, for which virial masses $\gsim10^{11}\,\mo$ are 
required on average for a halo to host stars. The maximum conversion efficiency 
is found for virial masses $\sim 10^{12.5}\,\mo$, about $\simeq 0.5$ dex 
larger than the analogous relation found in {\sc illustris} 
and inferred from abundance matching. Moreover, the 
peak of the relation appears to be shifting, albeit only slightly, towards large 
halo masses. At the high-mass end ($M_{200,{\rm crit}}\sim 10^{13.5}\,\mo$), 
the discrepancy among curves gets smaller, but the main trend with the seed 
magnetic field intensity is still present.

\section{Summary and conclusions}\label{sec:conclusions}
We have investigated how the presence of cosmological magnetic fields affects
the global properties of the galaxy population using cosmological MHD
simulations. Our main results can be summarized as follows. 

\begin{enumerate}
  \item For seed B fields $\gsim10^{-9}\,\G$, the increased magnetic pressure 
  hampers gas accretion into low-mass dark matter haloes. This causes a reduction
  of the cosmic SFRD, which becomes more marked for increasing seed field intensities.
  \item The suppression of the cosmic SFRD also results in a reduction of the 
  \rev{total galaxy number at all redshifts and of present-day galaxy stellar
  mass function in all stellar mass bins} and in particular for $M_\star \lsim 
  10^{10}\,\mo$. Again, this effect is stronger for larger seed B  
  fields.
  \item At fixed virial mass, haloes tend to contain significantly less stars for
  increasing seed field intensities. This is particularly evident for the less 
  massive haloes. Indeed, for the largest seed field considered in the simulations,
  virial masses $\gsim10^{11}\,\mo$ are required for a halo to host stars.
\end{enumerate}

Our simulations neglect any change that the presence of a seed magnetic field can 
induce on the initial matter power spectrum. Several authors 
\citep[e.g.][]{Sethi2005, Sethi2009, Yamazaki2006, Fedeli2012} have argued that 
seed magnetic fields can affect the distribution of power especially at 
small/intermediate scales, at which the clustering of the matter would be 
enhanced. By not considering these variations to the matter power spectrum, 
we might overestimate the effect of large B seed field on small haloes in our runs. 
Moreover, our cosmological volume is fully magnetized at the initial redshift, 
which again may reinforce the effects of B fields on the
gas with respect to a scenario in which the field is first
generated within galaxies and then ejected in the intergalactic space.

Another point to bear in mind is that degree of reduction of the stellar content 
and star formation in haloes might also depend on the choice of the feedback 
model adopted in the simulations. The fiducial {\sc illustris} feedback model 
used in our runs is at the upper end of what is energetically plausible, in 
particular for the generation of galactic winds. Therefore, it is conceivable 
that (strong) seed magnetic fields would cause a smaller suppression of the star 
formation rate and galaxy stellar masses with less efficient feedback loops, 
helping to alleviate some of the tensions with the observations that we have
discussed.

Even with these caveats, from our runs emerges a consistent picture in 
which primordial magnetic seed fields and galaxy properties are closely related. 
In particular, primordial seed magnetic fields compatible with the currently 
available upper limits leave detectable signatures on global properties of the 
galaxy population, which can potentially be exploited, in conjunction with the 
existing methods, to put tighter constraints on their intensities.

\section*{Acknowledgements}
We thank an anonymous referee for a constructive report and Lars Herniquist, 
Philip Mocz, Ruediger Pakmor, David Spergel, Volker Springel and Paul Torrey for 
their insightful comments. We further thank Volker Springel for giving us access 
to \arepo. MV acknowledges support through an MIT RSC award. The 
simulations were performed on the joint MIT-Harvard computing cluster supported 
by MKI and FAS. All the figures were created with {\sc matplotlib}
\citep{matplotlib}. 

\bibliographystyle{mnras}
\bibliography{letter}

\label{lastpage}
    
\end{document}